\documentstyle[11pt,newpasp,twoside,epsf]{article}
\def\edcomment#1{\iffalse\marginpar{\raggedright\sl#1\/}\else\relax\fi}
\reversemarginpar

\def\Msun{\,M$_\odot$}
\def\Rsun{\,R$_\odot$}

\begin{document}
\title{The Formation of Black--Hole X--Ray Transients}
 \author{Vassiliki Kalogera}
\affil{Harvard--Smithsonian Center for Astrophysics, 60 Garden St.,
Cambridge, MA 02138}

\begin{abstract}
 Studies of the observed characteristics of black--hole (BH) X--ray
binaries can be provide us with valuable information about the process of
BH formation. In this paper I address some of the aspects of our current
understanding of BH formation in binaries and point out some of the
existing problems of current theoretical models. In particular, the
measured orbital periods and donor--star properties indicate that a
common--envelope phase appears to be a necessary ingredient of the
evolutionary history of observed BH X--ray transients, and that it must be
associated only with a modest orbital contraction. The timing of this
common--envelope phase is crucial in determining the final BH masses and
current evolutionary models of mass--losing massive stars place strong
constraints on the possible masses for immediate BH progenitors and wind
mass loss from helium stars.  Last, it is interesting that, even in the
absence of any source of mass loss, the highest helium--star masses
predicted by current evolutionary models are still not high enough to
account for the measured BH mass in V404 Cyg ($> 10$\,\Msun ). An
alternative for the formation of relatively massive BH may be provided by
the evolutionary sequence proposed by Eggleton \& Verbunt (1986), which
invokes hierarchical triples as progenitors of BH X--ray binaries with
low--mass companions.

 \end{abstract}

\section{Introduction} 

Radial velocity measurements of the non--degenerate donors in X--ray
transients at quiescence combined with information about donor spectra and
optical light curves allow us to measure the masses of accreting compact
objects (e.g., Charles 1998), as well as other binary properties (e.g.,
orbital periods, donor masses and spectral types, kinematic properties).
At present, measured masses for nine X--ray transients exceed the optimum
maximum neutron mass of $3$\,\Msun (e.g., Kalogera \& Baym 1996) and the
binaries are thought to harbor black holes (BH). These observed BH X--ray
transients seem to form a rather homogeneous sample and studies of its
properties as a whole can shed a light to their evolutionary history and
the process of BH formation.

Black--hole X--ray transients are similar to low--mass X--ray binaries
with neutron stars in that mass transfer is driven by Roche--lobe overflow
and the donors are less massive than the BH. This maximum mass ratio of
about unity (see Kalogera \& Webbink 1996) allows the donor to transfer
mass stably to the compact object. However, for the majority of the BH
X--ray transients (including six BH candidate systems based on their
spectral properties, see Chen, Shrader, \& Livio 1997) the donors are less
massive than $\sim 1$\Msun, much less massive than the typical BH masses.
Only two systems, J1655-40 and 4U1543-47, have intermediate--mass donors
(more massive than $\simeq 2$\,\Msun) of $\sim 2.3$\Msun\, and $\sim
2-3$\Msun, respectively\footnote{One more X-ray transient, V4641 Sgr, with
an intermediate--mass donor of 5--8\,\Msun\, has been reported since the
time this talk was presented; Orosz et al.\ (2000).}. Further, orbital
periods are found in the range of a few hours to days, much smaller than
orbital periods of binary progenitors that could accommodate the radial
expansion of evolved massive stars (BH progenitors).

In what follows, we consider a number of the observed characteristics of
BH X--ray transients and use them as clues and windows to the evolutionary
history of these binaries, BH formation, and helium--star evolution. 

\section{Binary Orbital Periods and Common--Envelope Evolution} 

Currently observed orbital periods vary from just a few hours to a couple
of days typically (with the exception of V404 Cyg, which has an orbital
period of about 6 days). These orbital periods along with the measured BH
masses and the estimated donor masses (based on spectral type
classifications) clearly indicate that the orbital separations of the
X-ray transients are of order $10$\,\Rsun (even for V404 Cyg where the
donor is on the giant branch, we derive $\simeq 35$\,\Rsun). Since evolved
massive stars typically expand to $\sim 1000$\,\Rsun (even for stars that
lose their hydrogen envelopes though winds, see Schaller et al.\ 1992),
the progenitors of the observed binaries must have experienced a drastic
orbital contraction. Based on this simple observation it is generally
accepted that a common-envelope phase is necessary to explain the current
short orbital periods.

We can look into this question in some more detail. Let us for a moment
assume that the primordial binary was wide enough to avoid
common--envelope evolution. Then the only possible way to achieve the
required orbital contraction is through natal kicks imparted to black
holes. The magnitude of these kicks may be quite modest, since the effects
of kicks on binary characteristics are more important when their magnitude
is comparable to orbital velocity of the binary. For massive binaries with
orbital separations well in excess of $1000$\,\Rsun , orbital velocities
lie in the range $\simeq 10-50$\,km\,s$^{-1}$. However, the required
degree of orbital contraction is rather large, with a ratio of
circularized post--supernova to pre--supernova orbital separations: 
 \begin{equation}
 \alpha_C\equiv \frac{A_C}{A_{\rm in}} < \frac{10}{1000}. 
 \end{equation}
 Under the assumption of angular momentum conservation during
circularization, it is: 
 \begin{equation}
 \alpha_C=\alpha\,(1-e^2),
 \end{equation}
 where $\alpha\equiv A/A_{\rm in} $, $A$ is the immediate
(non--circularized) post--supernova separation, and $e$ is the
post--supernova eccentricity. Further, a strict lower limit on $\alpha$
can be derived analytically (e.g., Flannery \& van den Heuvel 1975;
Kalogera 1996):
 \begin{equation}
 \alpha\equiv\frac{A}{A_{\rm in}} > \frac{1}{2}. 
 \end{equation}
 Combining equations (1), (2), and (3) we derive that the required degree
of post--supernova orbital contraction can occur only for binaries with
post--supernova eccentricities in excess of 0.99\footnote{For V404 Cyg,
the corresponding lower limit on eccentricity is 0.96.} (!). Analysis of
post--supernova binary characteristics, for cases of significant mass loss
and kicks comparable to the pre--supernova orbital velocity shows that
only a negligible fraction (less than 0.1\%) of surviving (bound) systems
acquire eccentricities that high (see Figures 4 and 10 in Kalogera 1996).
Such a low survival probability in the absence of a common--envelope phase
would lead to a negligible formation rate for BH X--ray transients (see
also Portegies--Zwart, Verbunt, \& Ergma 1997). Therefore, we conclude
that a common--envelope phase is {\em necessary} for the formation of all
of the observed BH X--ray transients.

\section{Evolutionary Constraints}

Given that the progenitors of BH binaries must experience a
common--envelope (CE) phase, which leads to both orbital contraction and
loss of hydrogen envelope of the BH progenitor, it is quite possible that
the evolutionary history is very similar to that of low--mass X--ray
binaries with neutron stars (e.g., van den Heuvel 1983). In what follows,
we consider structural and evolutionary requirements for such a formation
path. In the next section we examine how these requirements determine the
donor properties of the BH binaries formed (for more details, see Kalogera
1999). 

The binary primary must be massive enough so that its helium core exposed
at the end of the CE phase collapses into a BH. The X--ray phase is
initiated when the donor fills its Roche lobe because of orbital shrinkage
through magnetic braking (for low--mass donors) or of radial expansion
through nuclear evolution on the main sequence (for intermediate--mass
donors).

Black--hole binary progenitors evolve through this path provided that the
following constraints are satisfied: 
 \begin{itemize}
 \item
 The orbit is small enough that the primary fills its Roche lobe and the
binary enters a CE phase. 
 \item
 At the end of the CE phase the orbit is wide enough so that both the
helium--rich primary and its companion fit within their Roche lobes. The
constraint for the companion turns out to be stricter.
 \item
 The system remains bound after the collapse of the helium star. In the
case of small or zero kicks imparted to the BH, this sets an upper limit
on the mass of the BH progenitor.
 \item
 After the collapse, the orbit must be small enough so that mass transfer
from the donor starts before it leaves the main sequence and within
10$^{10}$\,yr. 
 \item
 Mass transfer from the donor proceeds stably and at sub-Eddington
rates. This sets an upper limit to the donor mass on the zero-age main
sequence and to the orbital size for more evolved donors.
 \end{itemize}

\section{Donor Masses in Black--Hole X--ray Binaries}

For a specific value of the BH mass, the above constraints translate into
limits on the properties, circularized post--collapse orbital sizes
($A_C$)  and donor masses ($M_d$), of BH binaries with Roche--lobe filling
donors. The relative positions of these limits on the $A_C-M_d$ plane and
the resulting allowed $M_d$ ranges are exactly determined by three well
constrained model parameters:
 \begin{figure}
 \plotfiddle{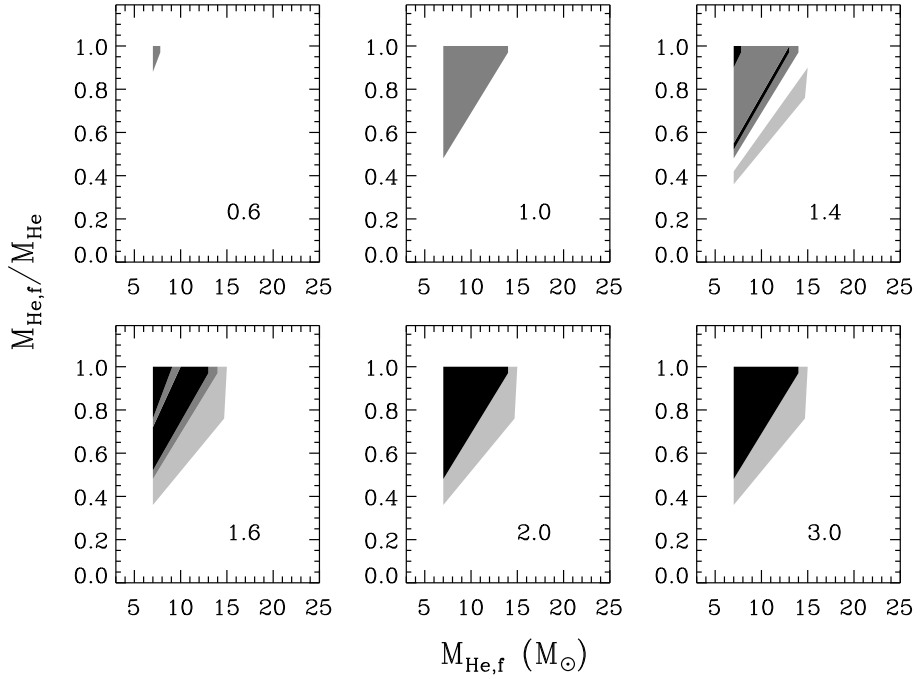}{3.5in}{-270.0}{80}{80}{320}{-100}
 \caption{Limits on the parameter space of the final (pre--collapse)
helium--star mass, $M_{He,f}$, and the ratio, $M_{He,f}/M_{He}$, for six
values of the $\alpha_{\rm CE}=$0.6, 1.0, 1.4, 1.6, 2.0, 3.0, and for a
7\Msun\, BH. Conditions in the unshaded areas do not allow the formation 
of BH binaries with main--sequence Roche--lobe filling donors; conditions
in the light--gray, dark--gray, and black areas allow the formation of
systems with only low--mass, only intermediate--mass, and both types of
donors, respectively.}
 \end{figure}

 \begin{itemize}
 \item 
 The amount of mass loss from the binary during BH formation,
characterized by the ratio $M_{He,f}/M_{\rm BH}$, where $M_{He,f}$ is the
mass of the helium--rich BH progenitor at the time of the collapse. For
the post--collapse system to remain bound it must be $1 \leq
M_{He,f}/M_{\rm BH} \leq 3$.
 \item
 The amount of mass lost in the helium--star wind between the end of the
CE phase and the BH formation, characterized by the ratio
$M_{He,f}/M_{He}$, where $M_{He}$ is the initial helium--star mass (at the
end of the CE phase). This ratio must lie in the range $0-1$.
 \item
 The CE efficiency, $\alpha_{CE}$, defined as the ratio of the CE binding
energy to the orbital energy released during the spiral--in of the
companion. Although the absolute normalization of $\alpha_{CE}$ is not
well determined (see Kalogera 1999), values higher than unity imply the
existence of energy sources other than the orbit (ionization or nuclear
burning energy).
 \end{itemize}

Note that the last two of the evolutionary constraints (\S\,3) depend {\em
only} on the BH mass, while $\alpha_{CE}$ affects {only} the upper limit
on $A_C$ (first of the constraints \S\,3). For different values of these
three parameters, the positions of the limits on the $A_C-M_d$ plane
change and three different outcomes with respect to the donor masses are
possible: BH binaries can be formed with (i) {\em only} low--mass; (ii)
{\em only} intermediate--mass; (iii) {\em both} low-- and
intermediate--mass donors.

The donor types as a function of the three parameters, $M_{He,f}$,
$M_{He,f}/M_{He}$, and $\alpha_{\rm CE}$, are shown in Fig.~1, for a
7\Msun\, BH. For $\alpha_{\rm CE}$ smaller than $\sim 0.5$, the orbital
contraction is so high that the donor stars cannot fit in the post--CE
orbits, and hence no BH X-ray binaries are formed. As $\alpha_{\rm CE}$
increases, CE ejection without the need of strong orbital contraction
becomes possible for the more massive of the donors, while formation of
binaries with low--mass donors occurs only if $\alpha_{\rm CE}>1.5$. The
results become independent of $\alpha_{\rm CE}$ for values in excess of
$\sim 2$, when the upper limit for CE evolution (first of the
constraints) lies at high enough values of $A_C$ that it never interferes
with the other limits.

The dependence of these results on the two mass--loss parameters (wind and
collapse) are determined by their association with orbital expansion. For
strong helium--star wind mass loss ($M_{He,f}/M_{He}<0.35$), the
progenitor orbits expand so much that donors less massive than the BH can
never fill their Roche lobes on the main sequence. Both low-- and
intermediate--mass donors are formed only if less than 50\% of the initial
helium-star mass is lost in the wind. Mass loss at BH formation is limited
to BH progenitors less massive than about twice the BH mass so that
post--collapse systems with low--mass donors remain bound. Note that
amounts of mass lost in helium-star winds and in BH formation are actually
anti-correlated. If one is close to the maximum allowed then the other
must be minimal (see Fig.\ 1).

\begin{figure}
 \plotfiddle{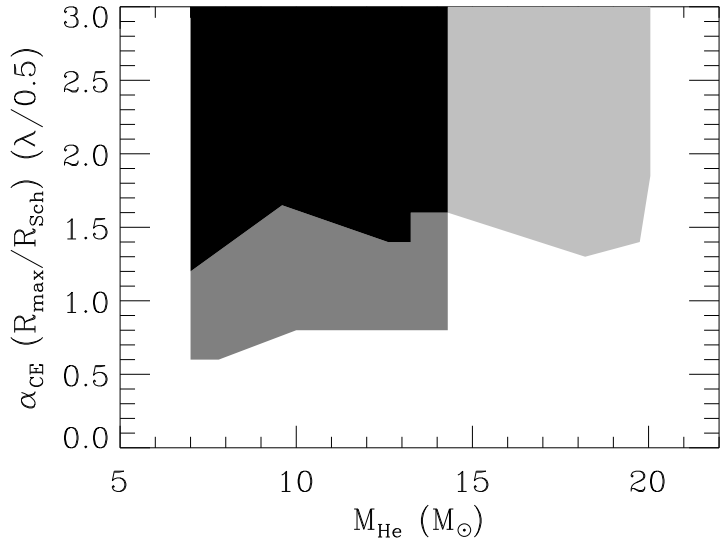}{2.5in}{-270.0}{55}{55}{230}{-60}
 \caption[]{Limits on the parameter space of the initial (post--CE)
helium--star mass, $M_{He}$, and the common--envelope efficiency,
$\alpha_{\rm CE}$, properly normalized (by the maximum stellar radii of   
massive stars (Schaller et al.\ 1992) and the central--concentration 
parameter, $\lambda$), for a 7\Msun\, BH. Shade coding is as in Figure 1.
}
 \end{figure}

The dependence on $M_{He,f}$ of the orbital expansion during helium--star
wind mass loss and BH formation is such that the ratio of circularized
post--collapse over post--CE orbital separations becomes independent of
$M_{He,f}$. This means that, for a specific BH mass, the position of the
limits on the $A_C-M_d$ plane depend only on the initial helium--star mass
and the CE efficiency. Indeed, in Fig.~1, the change of donor types
occurs along straight lines in the $M_{He,f}/M_{He}$--$M_{He,f}$ plane,
or else along lines of constant $M_{He}$. This simplifying property
allows us to combine the panels in Fig.~1 into one plot (Fig.~2). It is
evident that formation of 7\Msun\, BH X--ray binaries with both low-- and
intermediate--mass donors (as required by the observed sample) constrains
the common--envelope efficiency to relatively high values and the initial
helium-star progenitors at most twice as massive as the BH
(corresponding to initial primaries in the range 25--45\Msun).

Additional constraints can be obtained by examining the relative numbers
of systems with low-- and intermediate--mass donors formed for the
parameters in the black-shaded areas in Figs.\ 1 and 2.  The lifetimes for
the two different types are determined by the process that drives mass
transfer. The magnetic--braking time scale, for low--mass donors is
comparable to the nuclear evolution time scale of intermediate--mass stars
(Kalogera 1999). The number ratio then becomes equal to the ratio of birth
rates. The latter can be calculated using the derived limits on $A_C$ and
$M_d$ and assumed distributions of mass ratios and orbital separations of
primordial binaries. The results indicate that even when low--mass
companions in primordial binaries are strongly favored, BH binaries with
intermediate--mass donors are much more easily formed because of the
larger range of orbital separations allowed to their progenitors (see
Figure 3). Models predict a small fraction of intermediate--mass donors
(as seen in the current observed sample)  only for rather high
$\alpha_{\rm CE}$ values ($>3$) or for moderate (but still higher than
unity)  $\alpha_{\rm CE}$ values ($1.5-2$) and BH progenitors either
slightly more massive or twice as massive as the BH.

We note that these results are quite robust and do not depend on the
assumed BH mass nor the properties of primordial binaries (see Kalogera
1999).

\section{Stellar Evolution Models and Common--Envelope Phase} 

More careful consideration of the above analysis shows that the
constraints derived for helium--star wind mass loss depend on the timing
of the CE phase with respect to the evolution of the massive BH
progenitor. Current single--star evolutionary models (Schaller et al.\
1992; Hurley, Pols, \& Tout 2000) for massive stars ($>10$\,\Msun) of
solar metallicity with wind mass loss imply that a Roche--lobe overflow
(and hence a CE phase) can occur only either {\em before} or {\em after}
the star's core helium burning phase. During this phase wind mass loss
accelerates, leading to significant orbital expansion (Jeans mode of mass
loss), while the stellar radius remains almost constant. Also, for stars
more massive than about 25\,\Msun\, (about 35\,\Msun\, for Z=0.002), the
evolutionary models show a radial contraction until the star reaches core
collapse (see also Kalogera 1999).

\begin{figure}
 \plotfiddle{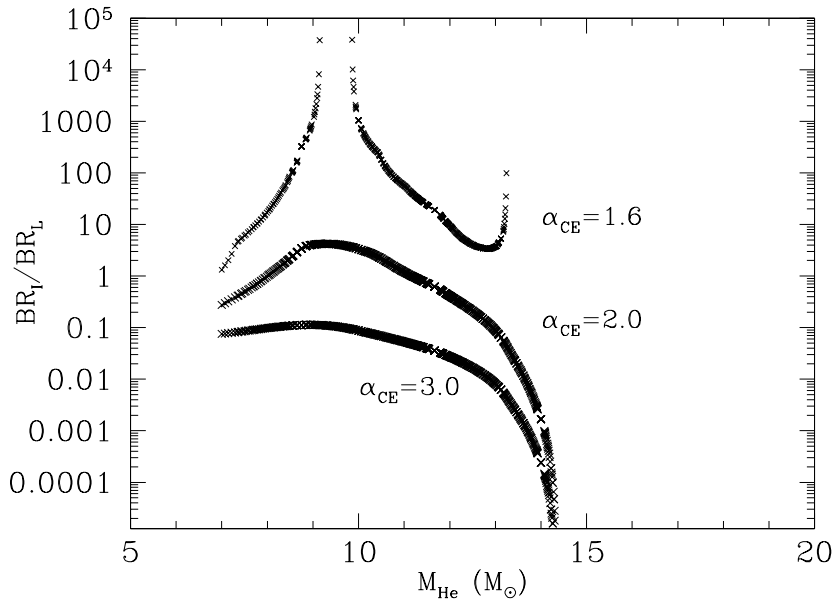}{2.8in}{-90}{50}{50}{-190}{265} 
 \caption{Ratios of the birth rate of BH binaries with intermediate--mass
donors to that of systems with low--mass donors as a function of the   
initial (post-CE) helium star mass, for three different values of
$\alpha_{\rm CE}$ properly normalized.  The mass-ratio distribution is
assumed to strongly peak at low values and therefore the ratios shown
could be regarded as lower limits.  }
 \end{figure}

Based on the above considerations, BH progenitors can fill their Roche
lobe only {\em before} core helium ignition (Case B mass transfer). The
exposed helium cores evolve through the complete core helium burning phase
before they reach core collapse. Taking into account this implicit
constraint, we are able to derive the associated constraints on
helium--star wind mass loss. Current models of helium--star evolution
through core helium burning (Woosley 1995) predict amounts of mass lost in
the wind significantly larger than the maxima allowed for BH X-ray
transient formation ($<50$\%). In fact, the final helium-star masses in
these models are $\sim 4$\Msun, far too small to explain the BH mass
measurements. Therefore, if the CE phase is initiated early in the core
helium burning phase of the primary, then helium-star winds must be much
weaker than thought until now. It is worth noting that more recent
empirical estimates of wind mass loss rates show a downward trend (Hamann
\& Koesterke 1998).

However, if the models for massive star evolution with winds do not
represent reality, and radial expansion (and hence Roche--lobe overflow
and CE evolution) is possible after the end of core helium burning, then
it is possible that the progenitors of BH binaries evolve through Case C
mass transfer (see Wellstein \& Langer 1999). In this case, the strength
of helium--star winds becomes irrelevant to the process of BH X--ray
binary formation. The reason is that the helium core is exposed only
through its core carbon and later burning phases, and the total duration
of these phases is so short that the wind mass loss is insignificant and
the helium--star mass remains essentially constant (Woosley 1995).

Given the uncertainties in models of massive star evolution and the
calculation of stellar radii, such a modification cannot be excluded at
present. In any case, it becomes clear that the existence of BH X--ray
transients requires that either the hydrogen--rich massive star models or
the strength of helium--star winds be modified.

\section{Discussion}

We have shown that the observed properties of the current sample of BH
X--ray transients provide us with clues to their evolutionary history and
BH formation. Specifically, we find that (i) common--envelope evolution is
necessary to account for the present tight orbits, (ii) orbital
contraction during the common--envelope phase must be moderate and
therefore CE efficiencies must be relatively high\footnote{Moderate
orbital contraction could also be achieved if the hydrogen envelope masses
are reduced through binary--enhanced wind mass loss as proposed by P.P.\
Eggleton.} (depending on the exact radii of massive stars and their
density profiles, significant contributions from energy sources other than
the orbit may be required), (iii) helium stars that form black holes are
at most twice more massive than the black holes at the time of collapse.
All these constraints do not depend on the details of the radial evolution
of massive stars.  We further find that current evolutionary models for
massive stars losing mass in winds appear to be in conflict with models of
wind mass loss from helium stars. Assuming that models massive star
evolution are more accurate than estimates of helium--star winds, we find
that wind mass--loss from helium stars must be limited so that these stars
lose at most half of their mass at the beginning of core helium burning.

Current stellar evolution models face one additional challenge posed by
the BH mass of V404 Cyg, if the upper end of the measured range is
confirmed ($10-14$\,\Msun; see, e.g., Bailyn et al.\ 1998). Even if we
ignore {\em any} mass loss from helium stars, predicted masses for helium
cores of massive stars cannot account for such a high BH mass (see
Wellstein \& Langer 1999; Hurley et al.\ 2000). 

An alternative model for the formation of BH X--ray binaries that invokes
the evolution of hierarchical triples has actually been suggested by
Eggleton \& Verbunt (1986). The basic idea is that the progenitor consists
of a high--mass inner binary with an outer low--mass companion. The inner
binary evolves to a high--mass X--ray binary, where Roche--lobe overflow
from the companion to the neutron star leads to a complete merger and
possibly the formation of a ``Thorne--Zytkow'' object. During the merger
the neutron star is expected to collapse into a BH, which can continue to
grow through accretion. If the massive envelope around the BH expands as
giants do then it is possible that this envelope will eventually engulf
the outer companion. The resulting spiral--in and envelope ejection would
then lead to the formation of a tight binary with a BH and a Roche--lobe
filling low--mass companion. 

Such an alternative formation path can very easily overcome difficulties
with (initial and final) helium--star masses and requirements for large
(possibly unphysical) common--envelope efficiencies. Furthermore, a
significant fraction of stars appear to be members of multiple systems.
These considerations could motivate a detailed analysis of this
triple--star formation path for BH binaries. Such an analysis would
eventually include the study of a number of very interesting problems:
orbital stability of the triple system throughout the long--term phases of
its evolution, the effects on the outer orbit of wind mass--loss and
supernova explosions occurring in the inner binary, the unsettled question
of the stability of ``Thorne--Zytkow'' objects (see, e.g., Cannon et al.\
1992, but also Fryer, Benz, \& Herant 1996) and their subsequent
evolution, the frequency and orbital properties of triple systems.

\bigskip\noindent{\bf Acknowledgments.}
 I would like to thank N.\ Langer, R.\ Taam, R.\ Webbink, and S.\
Wellstein for useful discussions. Support by the Smithsonian Astrophysical
Observatory through a Harvard-Smithsonian Center for Astrophysics
Postdoctoral Fellowship and a Clay Fellowship is also acknowledged.

\end{document}